# CROATIAN PUBLIC COMPANIES FOR ENERGY DISTRIBUTION AND SUPPLY: INTEGRATION OF INFORMATION SUBSYSTEMS

**[1]VLADIMIR SIMOVIC, [2]MATIJA VARGA, [3]VLADIMIR SIMOVIC**

[1,2]University North- Koprivnica Croatia,
[3]Faculty of Electrical Engineering and Computing University of Zagreb Croatia
E-mail: [1]vsimovic@unin.hr, [2]maavarga@gmail.com, [3]vsimovic@tvz.hr

**Abstract-** This research is about integration of information subsystems from:information system procurement, financial information system, information system security, technical information systems and legal information systems, and about their mutual dependence and close connections in Croatian public companies for energy distribution and supply. Also, herewe research the main goals of procurement information system which must be achieved in every organization because procurement process takes place in every public organization. Based on the model of the business technology matrix their processes can be executed by other companies engaged in similar activities. This research paper describes the timing of the sub processes, also. The timing of sub processes needs to be reduced as much as possible to achieve the planned results at the exit of the sub processes so that the costs of running sub-processes are equal to or lower than they had been so far, but with a higher quality output. We discuss possible threats to the information systems organization, and how to protect electronic information in the process of restoration of electronic data base for the financial information system. At the end the research paper we explain the advantages and disadvantages of cloud computing, information security in the event of possible applications of computing in the clouds and activities for technical information system in the flow diagram.

## I. INTRODUCTION

In this paper we present the preliminary concept of research the information subsystems and ERP systems that are present in almost all sectors of public companies and in public companies for the distribution and supply of gas.The main research topic:‘integration of information subsystems’ was modeled and developed on specific public companies work examples, wherewe show how information subsystems are working from technical, engineering and economic aspects.Consequently researchpaper is interesting source of research information for scientists from different branches of information sciencesEffective data management in the information systems and the newpossibilities of usage of business decisions opportunities developedwith information systems for decision support can support the top management in making important decisions. Decisions can be made in the information subsystem procurement on the basis of information from the legal information subsystem and financial information subsystem. Relevant information required in the procurement process can be collected from the security team of the information system and technical information subsystem. All valuable information on which key decisions are made by managers should be protected, whether they been in digital or analog form.

The research aimswere: (a) the model shown functioning information subsystem procurement, financial information subsystems, security policy information system, technical information subsystem and legal information subsystem; (b) describe their mutual interdependence and integration are closely observed within the organization; (c) show business system with reference to the information system of the company for the distribution and supply of natural energy source; (d) make an outline and document all the goals of the information subsystem in the technical sector and the sector of general and legal jobs; (e) detailed elaboration of models to display all business processes and flows of documents; (f) model display integration of information (sub)systems using basic parts information systems; (g) measure the time required for the process and time compared with the previous survey.

The company that processes observed is due to its specific business as it deals with distribution and supply of natural gas energy. The above information subsystems are essential for the efficient functioning of the entire organization regardless of the type of ownership. The main task of information subsystem procurement is to obtain all the information they need to procure resources and other funds for the work of the organizations at the appropriate criteria. Timely information at the appropriate time to enable the (pot) procurement processes generates profits. The tasks of the procurement process are: lower costs of obtaining resources and funding that enables efficient operation of the system and reduction of entropy. Today, the expansion of modern technology, there are potential threats directed at information systems organization, and describes how the protection of information in electronic form and methods of restoration of the database.

The integrated information system rationally and efficiently processed within a reasonable time accounting events referring to operations of assets, equity and liabilities. About the same changes in the financial position of the company management accounting prepares financial statements and informs the management company. At the same time in real-time links all users of an integrated information system with a unique database. (Mulahasanović, 2015).Within financial information subsystem used different methods and models to support decision-making. These information systems are presented as essential for the management of the company. information subsystem procurement is crucial for procurement of information and resources for smooth processes in organizations. Five stated information subsystems make information structure observed the organization in its entirety. Financial information subsystem is essential for recording business events and business changes and analysis of financial statements. Security information system is mainly focused in a way that protects the relevant, reliable, valuable information and the information transmitted in analog form that is relevant to the organization. Technical information subsystem of the organization is responsible for the display of real data from the process image of objective reality the management company from the field in moving calibration, distribution and maintenance. Technical information subsystem generates a realistic image data from objective reality based on processes performed in gauging office, on the field where it is the maintenance of the distribution system and where the viewing takes place preventive installation. Legal information subsystem enables greater security of information systems and it is in charge of Court Based Dispute Settlement instigated by: irresponsible business partners, representatives of households or internal company employees.

## II. THE IMPORTANCE OF INTEGRATION OF INFORMATION SUBSYSTEM

The aim of the information system is to provide business system with necessary and relevant information for the smooth execution of the process and system management. Procurement information subsystem in most cases is a complex system that enables: communication company with its customers and suppliers, good tracking flows of funds for the work, creating the conditions for monitoring the financial elements of business relationships, preparing and transmitting information in a financial information subsystem, specifically the process of accounting (Panian&Ćurko, 2010: 93).It is not wrong to point out the importance of the information subsystems procurement because of it begins execution of the process. Procurement information subsystem is used to collect information necessary for the smooth running of processes in organizations and public procurement funds for the work. Legal information subsystem in the company ensures that the decision-making process in public procurement is viewed.Fig.1 was made by authors and based: mainly on the scientific work: Šimović, V., Varga, M., Oreški, P. Case Study: an Information System Management Model. Management Information Systems. Vol. 7.(2012), No. 1, pp. 013-024 (Fig.1) and on basic logic of the Von Neumann original model on which today operateall modern computers and mobile devices.

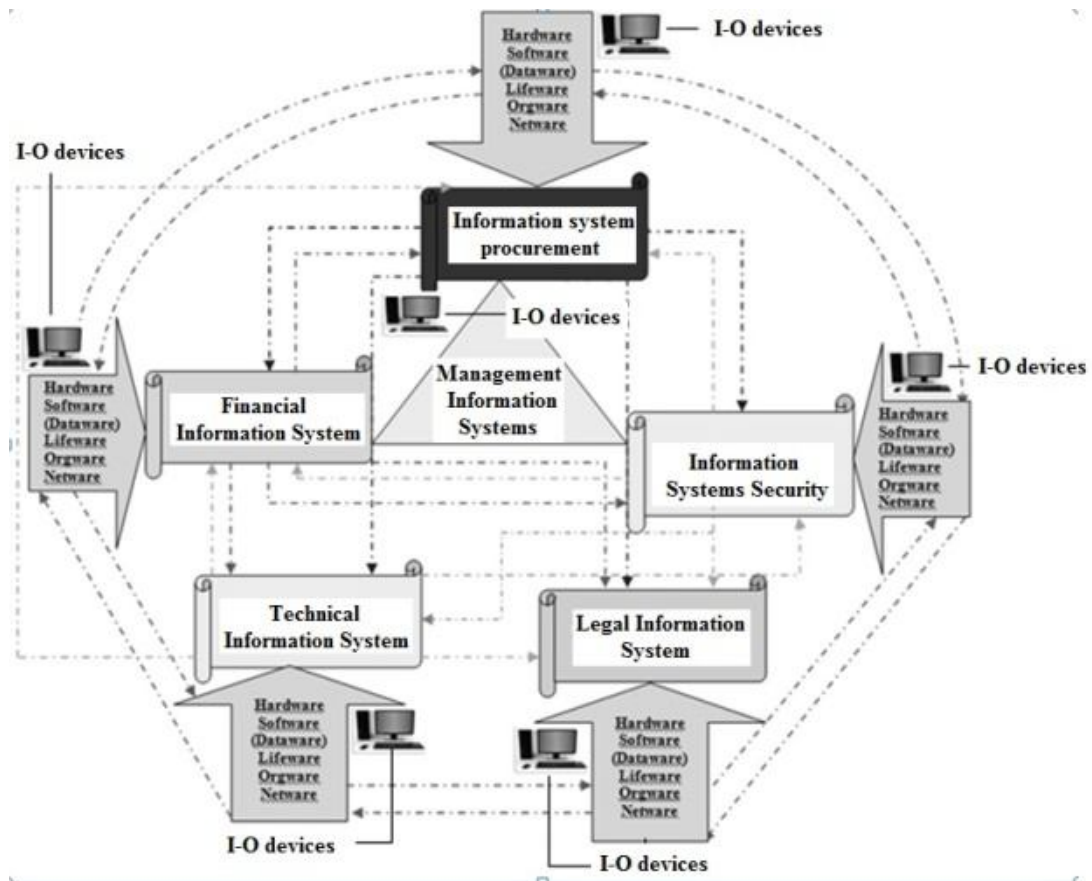

**Fig.1. Illustration of connection between elements of information subsystems and components of business information system**

The procurement process and its (sub) processes wasused for the procurement of information, goods, capital equipment, services and works. It is not wrong definition that the procurement generates profits (Ferišak, 2006). With the acquisition of more convenient means for working at a lower price than the real value is created and profits were also lower costs caused by procurement. It follows evidence of a close link information subsystem procurement, financial information subsystem, information system security, technical information subsystem and legal information subsystem (Fig.1). Model integration (Fig.1) shows the coordination of management information systems, and all components of the information system. Fig.1 shows that in each information (sub) system data can be entered by the user information system (Lifeware) through UI widgets that turn track (picture from the real reality) in a digital form using (IS) computers and data stored in the database (Dataware). The data is based on a physical level also (simultaneously) written to the hardware of the computer (ie. the tanks for storage) in the form of binary records. Netware allows access to the data (Dateware-in) for any modules of the information system only to human resources (Lifeware-in) who have access rights granted by a user in charge of the security of information systems. Information systems that have the same integration of elements in Fig.1 provide benefits as ERP (ie. ERM and / or ERP II) systems, which are a direct successor MRP2 systems which are: increased productivity, reduced inventory costs, production and logistics, improving communication with customers , integrating enterprise functions into one unit, and lay the basis for supply chain and e-business (Vuković, A. – Džambas, I. – Blažević, D. 2007: 42). List of world's largest manufacturers of ERP (enterprise software) that provides these benefits and associated information subsystems with different departments are: SAP, Enel, DataLab, Infosistem, Microsoft, ORACLE, Sage, Omega Softver, IN2 (vidilab, 2009., p.:3, 8, 10, 12, 14, 16, 22).

Financial information subsystem records all business events occurred in the organization takes care for the availability of cash funds. Security of information systems can't function without a financial information subsystem and provides funding, legal information subsystem that it legally protects and ensures that decision-making process is clear and public, and technical information subsystem. Financial information subsystem can't function without ensuring the security of information systems and legal information subsystem that it constantly protects and no technical information subsystem that takes care of the smooth running of the main processes of supply and distribution of energy.The aim of legal information subsystem is to improve the (way) process and activities: (a) the legal field; (b) area of public procurement; (c) the financial area; (d) the area of security; and (e) the technical area.

Fig.1 shows the integration of the above information subsystems and integration of the information systems as part of the program, the hardware part, organizational unit, personnel section (part of the Human Resources Management), power supply and a data part. Presented parts information system necessary for the smooth running of business processes and to improve efficiency of business

## III. ANALYSIS: AIMS OF PROCUREMENT AND LEGAL INFORMATION SUBSYSTEMS

The breakdown of the objectives of the information subsystems procurement and objectives of the legal information subsystem is a breakdown of the objectives of the information sub-goals to specific order and appropriately following the rules decomposition which reads: "Every parent must have at least two children ".

Fig.2 shows the decomposition of the goals of the information subsystems procurement. The present objectives include your (sub) goals which can be seen from the model. (Under) objectives are: to collect information about the conditions of supply, to collect information on the best supplier, to collect information on the possible reduction of the costs of supply, to collect information on the costs of storage, collect information on how to lower costs of care, market research and collect data from the field on the basis of an appropriate sample, collect information on the optimal amount of ordering, to collect information on the manner and time of delivery, to collect information on the need for training of staff in the procurement process, to collect information on the level of risk of procurement.

Fig.3 shows the decomposition of the objectives of legal information subsystems which are: to gather information that will ensure the right of companies to collect information to the decision making process in public procurement was viewed, collected information to ensure the legal security of the company and to collect information on debtors for services offered to businesses . Another goal of the decomposition diagram (which shows the objectives of the legal information system) is closely associated with the procurement process as it unfolds public procurement of goods, works and services under Articles of the Law on public procurement of goods, works and services.

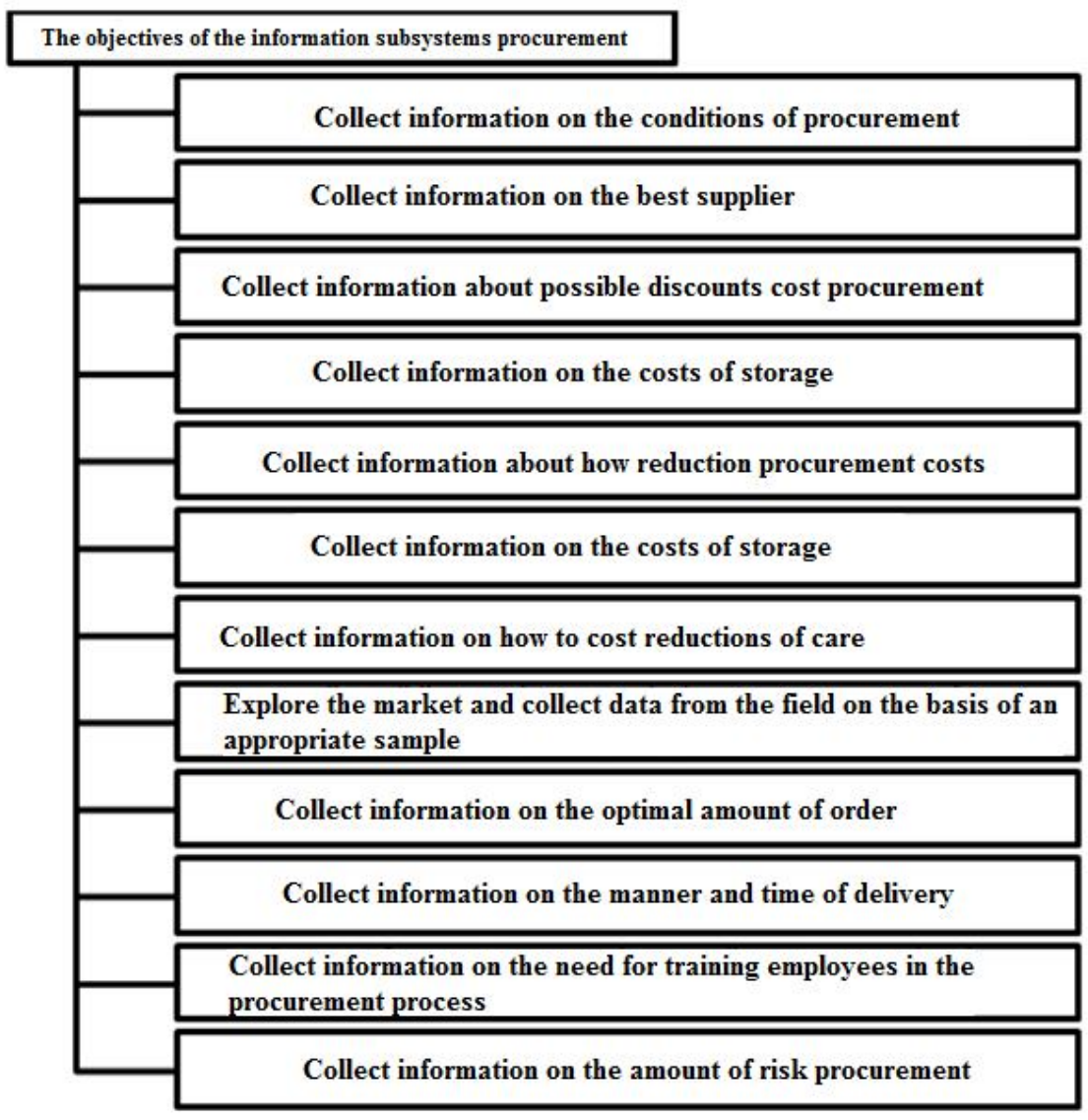

**Fig.2. Decomposition goals of the procurement information subsystems**

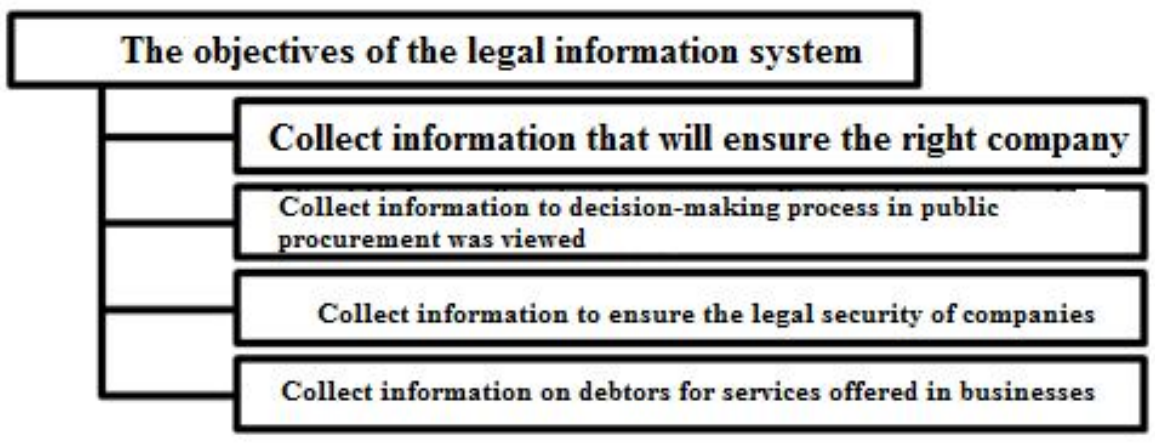

**Fig.3. Decomposition goals of the law information subsystems**

## IV. THE DATA CLASSES FOR INFORMATION SUBSYSTEMS OF THE PUBLIC COMPANIES FOR ENERGY DISTRIBUTION AND SUPPLY

Business technology matrix is often used in research for a clear overview of the integration of information subsystems. Business technology matrix is strictly defined two-dimensional mathematical structure over which they can implement formal operations such as checking the consistency of business technology optimization or organization, and describes the relationship between different factors (Brumec, 2008).Matrix processes range data and the method to display and process the data range and provides a clear picture of the business and technology of the system under its complexity of the system under examination has a scientific theory, the contribution system (Fig.4). The matrix is structured in such a way that there is no (way) process that generates only and does not use any class data. The process is a set of activities that occur in a specific order. Class information is logically designed and linked set of data, which refers to one guise or entity. Class data is in the process (or (sub) process) creates and uses. Business technology matrix for the observed information subsystems can be used when designing the data flow diagram.Fig.4. was based on business logic of the observed company, established 31 (way) process and 61 classes of data with tools for analytical data processing and on the basis of scientific articles (Šimović, V. - Varga, M. - Oreški, P. 2012: 16) and review article (Varga, 2011: 369).Fig.4 shows the detailed information of the public organization is divided into sections and its functioning.

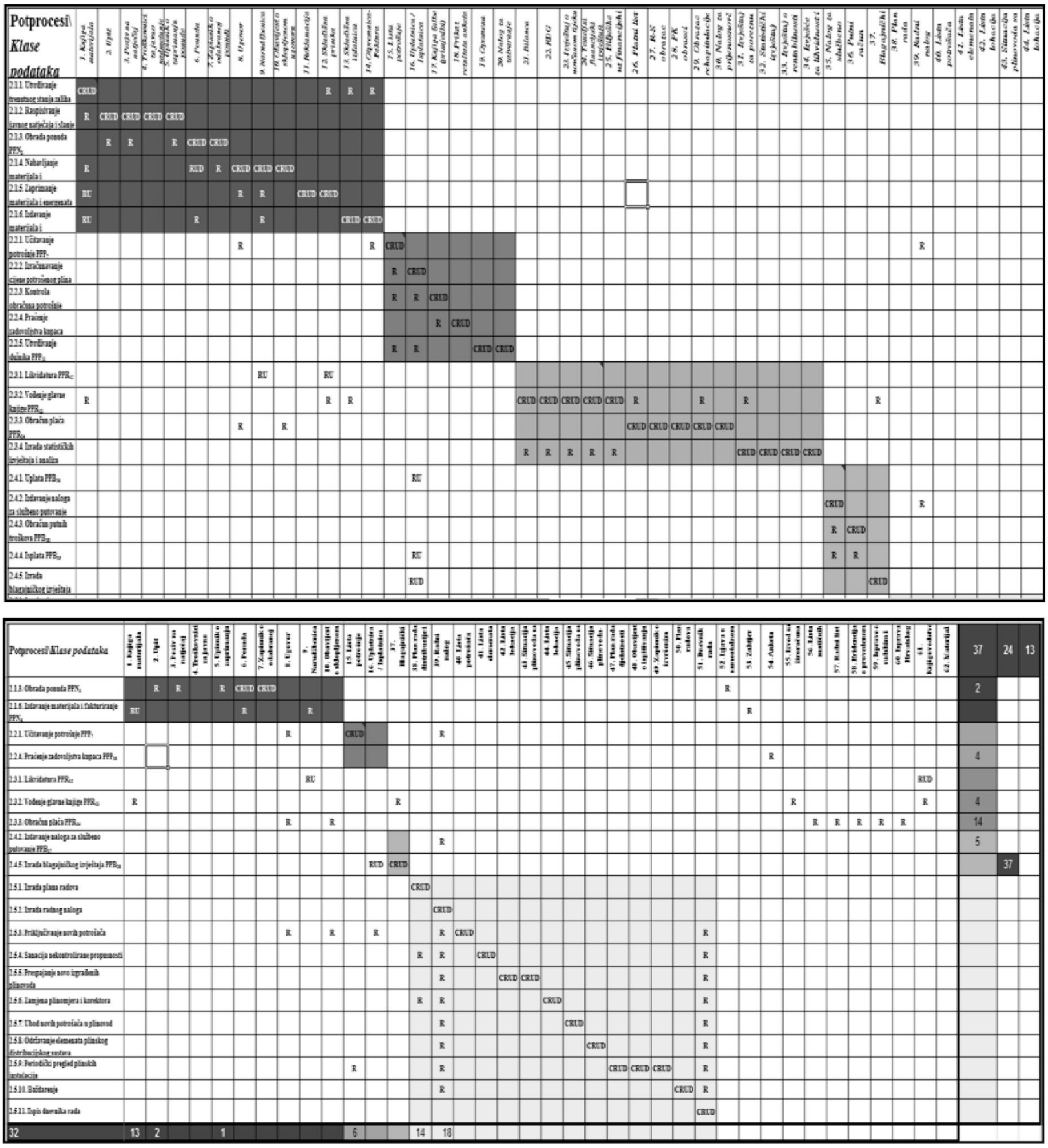

**Fig.4. Original integration matrix of information subsystems**

Business technology matrix is divided into two sections for easy reference.The matrix technology is hidden individual (sub) processes for detecting the integration of subsystems. Since the observed five information subsystems, is presented complex matrix of business technology companies for the supply and distribution of energy. Business technology matrix shows that the documents as information required information subsystem procurement (from suppliers to convince management of the organization that supplies that can independently achieve its objectives and carry out works), financial information subsystem, for ensuring the security of information systems, technical and legal information subsystem information subsystem.

To analyze the content of the matrix used is a tool for analytical data processing. The software tool is easier to quantify the how your matrix of business information technology sub-system procurement, a financial information subsystem, technical information and legal information subsystem contain sub-processes. Benefits matrix display technology business model of integration of information systems and processes and data classes are: Matrix provides a clear and transparent overview of all processes and data classes within the monitored information system as opposed to the flow chart of data that cannot be understood in view of large-scale information systems , the matrix shows how many times a particular process creates, reads, deletes and updates the specific class of data.

The duration of the individual processes, sub-processes and activities of the observed information system is shown in a separate table (Table 1).Technical information subsystem using class data: agreement, notice of conclusion of the contract, payment slips / paid in & paid list and spending and that of the class of data are created in the (sub) processes by the name of procuring materials and energy (contract, notice of conclusion of the contract), load consumption (list of consumption) and calculate the cost of consumed fuel (money order / paid in & paid).

## V. MEASURING THE SUCCESS OF THE INTEGRATED INFORMATION SUBSYSTEMS PROCESSES

Measuring the success of the process is done to control their execution. If the processes are performed in a way that gives the desired output (outputs) they do not need to improve, but with them to manage and control their performance. As arguments for the measurement process, PO Egnell (2001) lists the following reasons to work (Hernaus, T., 2006: 215): (a) knowledge about the success of the process; (b) affairs improvements focus on the right things; (c) comparison of goals and objectives of the process and the identification of deviations; (d) monitoring trends; (e) motivating employees to improve operations; and (f) eliminate those activities that do not add value.

All parts of the process should be measured in order to identify (sub) processes that do not create value; you need to carefully monitor such processes. The measurement process requires an understanding of the process and creates the ability to identify critical parts. Critical parts of the process can be measured when exactly determine what those (sub) processes, which significantly facilitates the way the management and control of business processes (Enstrom, 2002: 28-29). Efficient process management requires a wide range of measurement parameters. So they need the financial and non-financial data, both qualitative and quantitative (Kueng, 1999: 155). The process indicators can be classified into two categories: operational and structural.

Operating indicators measure how the process takes place over time. Specific indicators that can be used depends upon the specific process. Operating parameters are directly focused on the dynamic characteristics of business processes, structural indicators oriented static characteristics.They strongly influence the success of the process as it indirectly affects the dynamic characteristics. Three operational indicators are: the duration of the process, the bottleneck process and waiting times.The duration of the process is the average time between the end of the activities of successive units (Hernaus, 2006: 216).

| Potprocesi | 2.1.1. Utvrđivanje trenutnog stanja zaliha $PPN_1$ | 2.1.2. Raspisivanje javnog natječaja i slanje upita $PPN_2$ | 2.1.3. Obrada ponuda $PPN_3$ | 2.1.4. Nabavljanje materijala i energenata(robe, radova i usluga) $PPN_4$ | 2.1.5. Zaprimanje materijala i energenata $PPN_5$ | 2.1.6. Izdavanje materijala i fakturiranje $PPN_6$ | 2.2.1. Učitavanje potrošnje $PPP_7$ | 2.2.2. Izračunavanje cijene potrošenog plina za određeno područje $PPP_8$ | 2.2.3. Kontrola obračuna potrošnje $PPP_9$ | 2.2.4. Praćenje zadovoljstva kupaca $PPP_{10}$ | 2.2.5. Utvrđivanje dužnika $PPP_{11}$ | 2.3.1. Likvidatura $PPR_{12}$ | 2.3.2. Vođenje glavne knjige $PPR_{13}$ | 2.3.3. Obračun plaća $PPR_{14}$ | 2.3.4. Izrada statističkih izvještaja i analiza blagajničkog izvještaja $PPR_{15}$ | 2.4.1. Uplata $PPB_{16}$ | 2.4.2. Izdavanje naloga za službeno putovanje $PPB_{17}$ | 2.4.3. Obračun putnih troškova $PPB_{18}$ | 2.4.4. Isplata $PPB_{19}$ | 2.4.5. Izrada blagajničkog izvještaja $PPB_{20}$ |
|---|---|---|---|---|---|---|---|---|---|---|---|---|---|---|---|---|---|---|---|---|
| Vrijeme trajanja potprocesa (određeno na temelju podataka dobivenih od financijsko komercijalnog rukovoditelja(uzeti je prosjek)) Prije poboljšanja procesa | 0:02:00 | 0:50:00 | 0:60:00 | X | 0:30:00 | 0:30:00 | 0:04:00 po brojilu | 1:20:00 po očitavaču u prosjeku | X | 0:10:00 | 0:02:00 | 0:01:00 | X | 0:06:00 | X | 0:01:00 | 0:02:00 | 0:20:00 | 0:01:00 | 0:05:00 |
| Vrijeme trajanja potprocesa (određeno na temelju podataka dobivenih od financijsko komercijalnog rukovoditelja(uzeti je prosjek)) Nakon poboljšanja procesa | 0:01:00 | 0:55:00 | 0:55:00 | X | 0:20:00 | 0:25:00 | 0:02:00 po brojilu | 1:15:00 po očitavaču u prosjeku | X | 0:10:00 | 0:02:00 | 0:01:00 | X | 0:06:00 | X | 0:01:00 | 0:02:00 | 0:15:00 | 0:02:00 | 0:05:00 |
| Ljudski resursi promatrane tvrtke koji upravljaju potprocesima i korisnici su podataka | Skladištar, skladišni radnik, referenti nabave, financijsko-komercijalni rukovoditelj. | | | | | | Referenti prodaje, financijsko-komercijalni rukovoditelj. | | | | | Likvidator, voditelj računovodstva, glavni knjigovođa, knjigovođa. | | | | Operater obračuna plaća, blagajna, operater obračuna komunalnih usluga. | | | | |
| Objekt mjerenja | potprocesi | potprocesi | potprocesi | potprocesi | potprocesi | potprocesi | potprocesi | potprocesi | potprocesi | potprocesi | potprocesi | potprocesi | potprocesi | potprocesi | potprocesi | potprocesi | potprocesi | potprocesi | potprocesi | potprocesi |
| Učestalost mjerenja | ponavljajuća | ponavljajuća | ponavljajuća | ponavljajuća | ponavljajuća | ponavljajuća | ponavljajuća | ponavljajuća | ponavljajuća | ponavljajuća | ponavljajuća | ponavljajuća | ponavljajuća | ponavljajuća | ponavljajuća | ponavljajuća | ponavljajuća | ponavljajuća | ponavljajuća | ponavljajuća |

**Table 1:The duration of each process**

Table 1 shows the timing of the individual (sub) process in hh: mm: ss. The duration of individual (sub) process could not be determined, therefore, is placed unknowns X. Measurement duration and course of (sub) processes in the supply, financial and security system is significantly different than in the manufacturing information system. Table 1 shows that the operating parameters are applied, and that for each (sub) process in the process of a certain duration and the resources that take part in it. Based on Table 1 can be observed compare running times of individual (sub) process and it was found that for each (sub) processes need more time to run after a specified period has passed since the first measurement. Table 1 shows the human resources which are responsible for the running of (sub) process and have an effect on reducing the running times (the path) of the process and thus improve the performance of operation (path) of the process. If an observed process is not performing well or is too long run it is necessary to achieve the fundamental objectives to change the business processes which are less time, lower costs and higher quality of services. If the targets are achieved, the changes can be considered successful. Based on the timing of the process can be observed that in the second measurement processes shank took place and that they give the same results processing (as shown in Table 1).

## VI. FINANCIAL INFORMATION SUBSYSTEM INTEGRATED WITH CLOUD COMPUTING

The task of financial information subsystem is a recording of all business events in companies in the financial and value terms. Financial information subsystem and its modules for recording business events are similar in many organizations because the modules at the end of treatment processes must meet the set of rules that are defined in the articles of the Law on Accounting. Financial information subsystem of the software includes the following group of modules (Fig.5): General Ledger module, module analytical accounting (and other supporting books), which consists of (sub) modules such as accounting for fixed assets, the accounting of inventory, raw materials, calculation salaries and employee personnel records, (sub) module accounts receivable and suppliers, accounting for small inventory and spare parts.

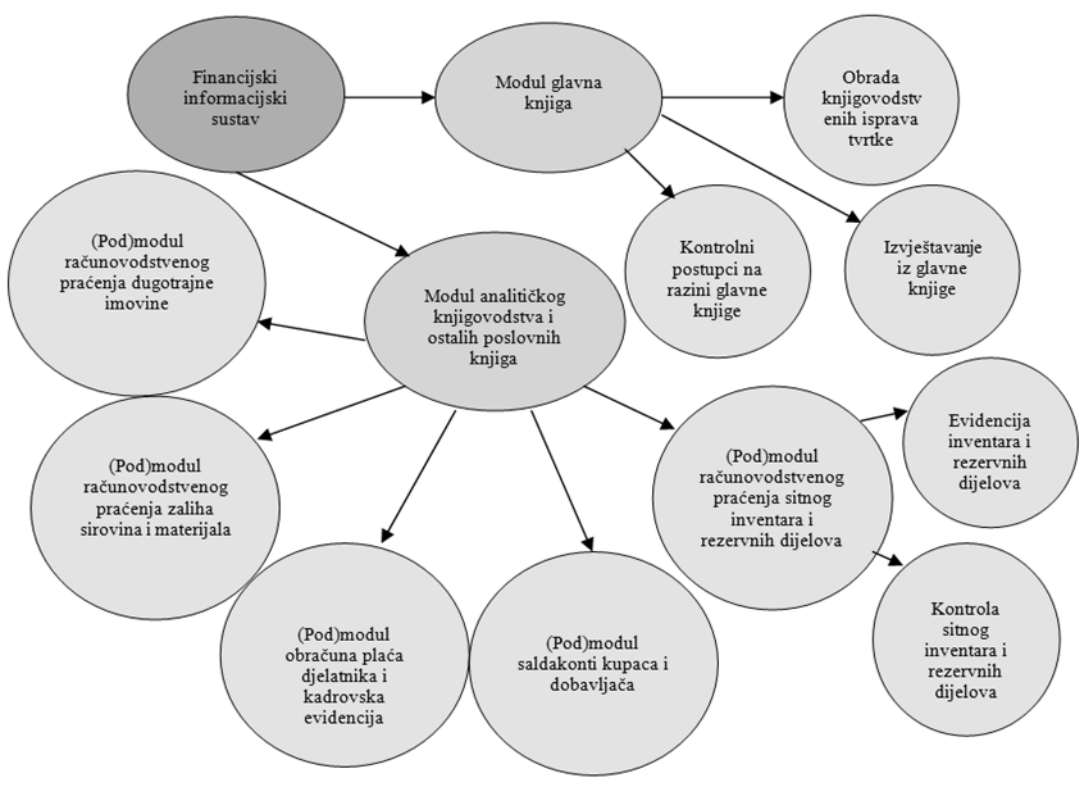

**Fig.5. Organization of financial information subsystem**
(Source: Based on the book: Panian, Ž. &Ćurko, K. (2010.) Business Information Systems. Zagreb, Element. p: 83.)

Fig.5 shows the organization of financial information subsystem. A feature module General Ledger Accounting in the process is that the data is recorded in the general ledger module relating to the last events occurred. The company may record events (financial changes) with cloud computing. Services cloud computing can create a company engaged in the development of cloud technologies. Companies for the development of cloud technology can enable the user to use the above modules. The content of the module is determined by the general ledger accounts and chart of accounts, which adheres to a business entity. In the general ledger module can be concluded that mandatory because its structure prescribed by the Law on Accounting. General Ledger module in most cases includes application reports (Panian&Ćurko, 2010: 84).

In the process of accounting documents inputs that create business events and transactions. The process of accounting has its output representing the result of the processing process. (Sub) processes within the accounting process are: liquidation, general ledger, and payroll, preparation of statistical reports and analysis of financial statements. (Sub) process liquidation includes the following activities: controlling forms of accounting documents, controlling text correctness and controlling the computational accuracy (financial value). Another name that is most commonly used for the (sub) process of liquidation was to control. After the completion of activities within the (sub) process of liquidation, the documents are entered into the general ledger. Activities (sub) process liquidation takes place employees in the workplace liquidator. Following the preparation of financial statements obtained certain information based on data contained in the general ledger. One of the tasks of the process of accounting the collection and processing of financial data from the financial statements and the presentation of information obtained the company's board, the supervisory board, the auditors, the owners of companies, trade unions, banks, public, suppliers, customers, employees and other interested people. The basic annual financial reports such as balance, RGD and additional data is sent to the environment is evident based on matrix technology business because of reports of interested buyers, suppliers, investors, institutions, government and others.

## VII. SECURITY POLICIES OF INTEGRATED INFORMATION SYSTEM

Security policy information system protects information systems within the organization, its processes and employees who participate in the execution process. Physical security is of great importance in the security information system organization. It is well known that the most common attacks on information systems caused by employees of a particular organization. Research performed and published in a book: D. Seger, K., VonStroch, W. „Computer Crime a Crimefighter's Handbook“, O'Reilly & Associates and prove that fact. The book states that the highest percentage of security problems caused by human error. Human error is most often occur due to lack of attention and lack of education.The second biggest cause of errors in information systems is a failure of hardware equipment, the third most common cause of employees who represent their position in the institution used for their own benefit, and the people in this way express their dissatisfaction with the company or supervisor (Kovacevic, 2015).

Table 2 shows the types of threats to the security of information systems. Table 2 presents the sources of threats, a description of the domain and the specific threat that shows specifically who can affect the security of information systems.In companies the user access to certain applications organized as follows: Head of Department in the coming new employee sends a request to open an account with the data on the level and access rights. In addition to other basic data, the application should include the job title, description of work and needed access rights. Once the user receives a password from the manager of the department can be changed so that he and the administrator can access specific data. Each employee must have a password to access a specific part of the application, i.e. particular module. Password changes usually every month, and if necessary, should be more than once a month. For a password would be desirable to include a combination of uppercase and lowercase letters and numbers. No way is not recommended to use the password given names, surnames, names of parents, children, birth date, name of the place of residence, street name and etc. For a password is not good to use the same set of characters and password must not write on paper and placed in a tray to avoid a third person came to the secret data and information of high value.

**Table 2:Types of security threats to information systems**

| Source of threats: | Description of the domain: | Concrete threats: |
|---|---|---|
| **Employees** | No respect for the internal security of company policy approved by the government.<br>Errors employees (intentional or not intentional) | Former employees<br>The current employees<br>Employees of the recruitment process<br>Since (potential) employees |
| **Processes and activities** | Absence of clearly defined regulations, business rules, processes, procedures.<br>Absence of well-defined sequence of unfolding functions, processes, sub-processes, activities.<br>Noncompliance procedures and regulations.<br>For too long execution processes, sub-processes and activities.<br>Do not conduct business process reengineering and sub that does not give the desired results at the right time when it is needed. | Employees<br>Customers<br>Suppliers<br>Service Providers<br>Business partners<br>Rivals<br>Media<br>Politicians |
| **Different systems** | Do not anticipated failure of hardware and software of the computer.<br>Not enough technical robustness of the system.<br>Not frequent replacement of old hardware parts of the computer. | Technical fault system within the prescribed use.<br>Technical defect in the system because of inadequate design or bad implementation of a new information system.<br>Technical failure of the gas distribution system within the building structure.<br>Infrequently create full, differential and incremental backups of data.<br>Do not take measures to protect the program.<br>Not setting limits access to certain information in the data bank.<br>viruses<br>Incident management information system.<br>No application of standards in the security system. |
| **External events (environmental impact)** | Natural disasters (lightning, snow, rain, flood, earthquake, dust storms, drought, heat, explosions, oil spills).<br>Disasters caused by human error.<br>Negligence foreign participants.<br>Legitimate actions external participants. | Natural events<br>Accidents<br>Malicious external actors<br>Negligent external actors<br>The conflict with the business interests of external actors. |

Source: Basis of the table is material: Klaić, A. (2010) Minimum security criteria and risk management of information security, URL:http://os2.zemris.fer.hr/ISMS/rizik/2010_klajic/SeminarskiRad_SRS_042010_AK.pdf, (6.8.2015)

The most common attack on password is the testing or guessing passwords. Testing or guessing passwords attack in which the offender is trying to access a particular system of random guessing passwords, while in most cases the method of trial and error. Although this attack seems a bit naive with some time it can be effective, especially when we know very well the person who set up the password.Taking into account the limitation of the number of attempts to access the computer system, the system must be adjusted in a way that limits the number of possible approaches. If a user attempts to more than three times to access the system with the wrong user name and password information system should reject it. A further possibility that it would be desirable to set that messages about the last access to e-mail is displayed ie. To display the last data access in the form of: the date, time and name of the Internet service provider from which the user has accessed your mail to be visible.

Cloud computing is not intended for the use of businesses small and medium enterprises, but cloud computing can use the large companies, companies that have more than 250 employees or companies that generate higher profits and the creation profits belong to the group of large companies. If a certain company has for each department made the appropriate applications gives it the ability to use cloud computing. It is recommended that departments of companies that do not have and do not manipulate the important data for the company as the use of cloud computing. When it comes to the security of electronic data and information with servers and cloud services technologies can talk about the shortcomings. Providers in the cloud enable data encryption, security, user names and passwords and rudimentary identity management. It is recommended that companies do not use cloud computing when it comes to recording business events within the financial information subsystem rather than being clearly defined rules of the contract of use. The issue of security is better mastered and there is no reason to believe that the same data would be more confident in their own storage systems databases. When a user preferred provider of cloud computing has more influence over the control of the information infrastructure provider that is dangerous for the user.

The user must comply with the regulations of the service provider to him the same would not be repealed account and access to data. It is possible to download the company that provides the service by other companies who decide on the continuation of cooperation with existing customers. The costs are a major reason why less use computing in the clouds. Sometimes they can be expensive to maintain applications in the cloud than on your own server or computer. The cloud computing usagecost can be advantage and disadvantage. It is therefore very important to consider the cost and return on investment. Most platforms in computing in the clouds is privately owned. If the user wants to move the system to another service provider, back in his own company may be a case of data loss or an increase in the cost of transmission of electronic data. Providers are trying as quickly as possible establish a set of standards in order to reduce the high costs. Openness and the use of cloud computing remains a risk. These are expected improvements in this area. A large number of service providers on the Internet do not provide a service level agreement. The trend of the service provider cloud computing is the inclusion of service level agreement, but at the same time shifting the costs of risk to service users.

## VIII. TECHNICAL INFORMATION SUBSYSTEM FOR DATA COLLECTION AND PROCESSING

The aim of the technical information subsystem is to provide a system and management companies all necessary and relevant information for the smooth execution of processes of distribution of energy and the smooth management of the system. Within the technical information system observed the company needed to collect information on the distribution of natural energy source that contain data on restoration, rehabilitation of damaged ports, rehabilitation uncontrolled bandwidth, connecting newly constructed lines, work at household connections, replacing measuring instruments and concealer, stab new consumer line. Intervention (Fig.7) includes work on removing anomalies that can occur at the consumer or to the distribution system, which significantly affected the safety and reliability of supply. Restoration of connections includes the planning, preparation, execution and monitoring of damaged cables and connections. In addition to collecting information on distribution of energy is necessary to collect information on the maintenance of installations. During the course of (sub) processes maintenance installation information is collected about uncontrolled bandwidth distribution system, maintenance reduction stations and distribution-reduction stations, maintenance of distribution shafts, maintaining condensing panes, keeping valves to block, detect underground cadaster.Technical information subsystem must contain the basis of electronic information on testing installations containing test data is not part of the installation of the measuring and testing of the measuring part of the installation. Fig.6 shows the decomposition of the goals of the technical information subsystem made according to the rules of analysis. According to the defined goals it can be perceived complexity of the technical information subsystem.

Fig.7 shows a flow chart of activities during the intervention. It is widely known that the diagrams the flow of activities developed Birtwistle in 1979 as a means of modeling. Interventions are necessary for the smooth running of the process of distribution. DiO is label for the distribution and maintenance, RN is the document - work order, while DR is mark indicates the flow chart daily work.Fig.7 shows the head of distribution and maintenance involved in making the plan work. The manager of distribution and maintenance creates work order and supervise the execution of work orders, decide on the continuation of the execution of work orders, work order control and daily work and take part in entering data into the tracking system. At the end of the archiving is performed work order and work diary. Emergency teams of outside intervention receive the work order and prepare the tools and materials necessary, execute work order and print the log. At the end of execution of the order and print the daily work compile the minutes.

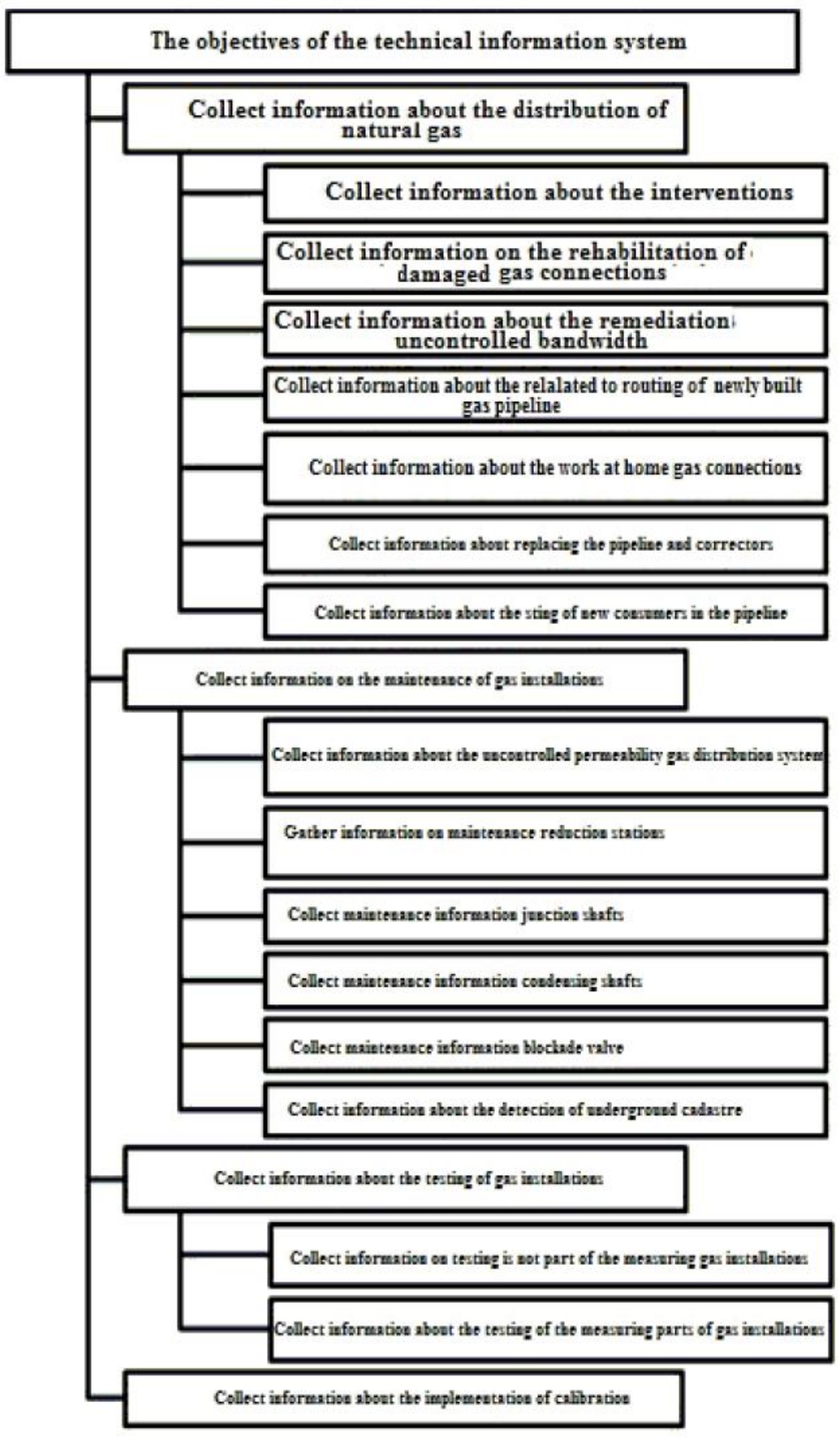

**Fig.6. The objectives of the technical information subsystem**

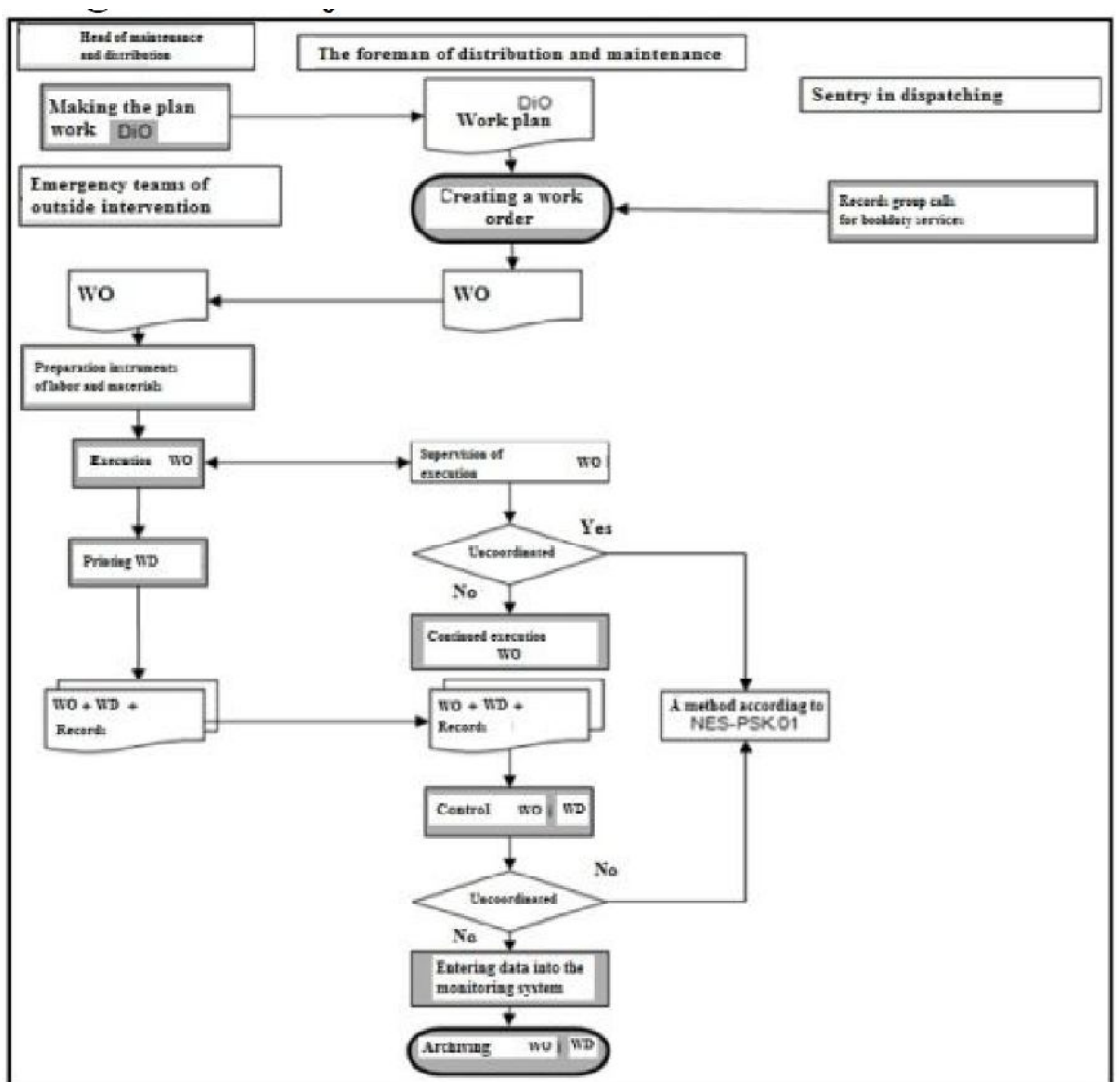

**Fig.7. A diagram shows the flow of activities at intervention**

(Source: Creating an author based on interviews of employees observed the company and manual quality (PQ 9001. Prema EN ISO 9001:2000). (12.2.2007))

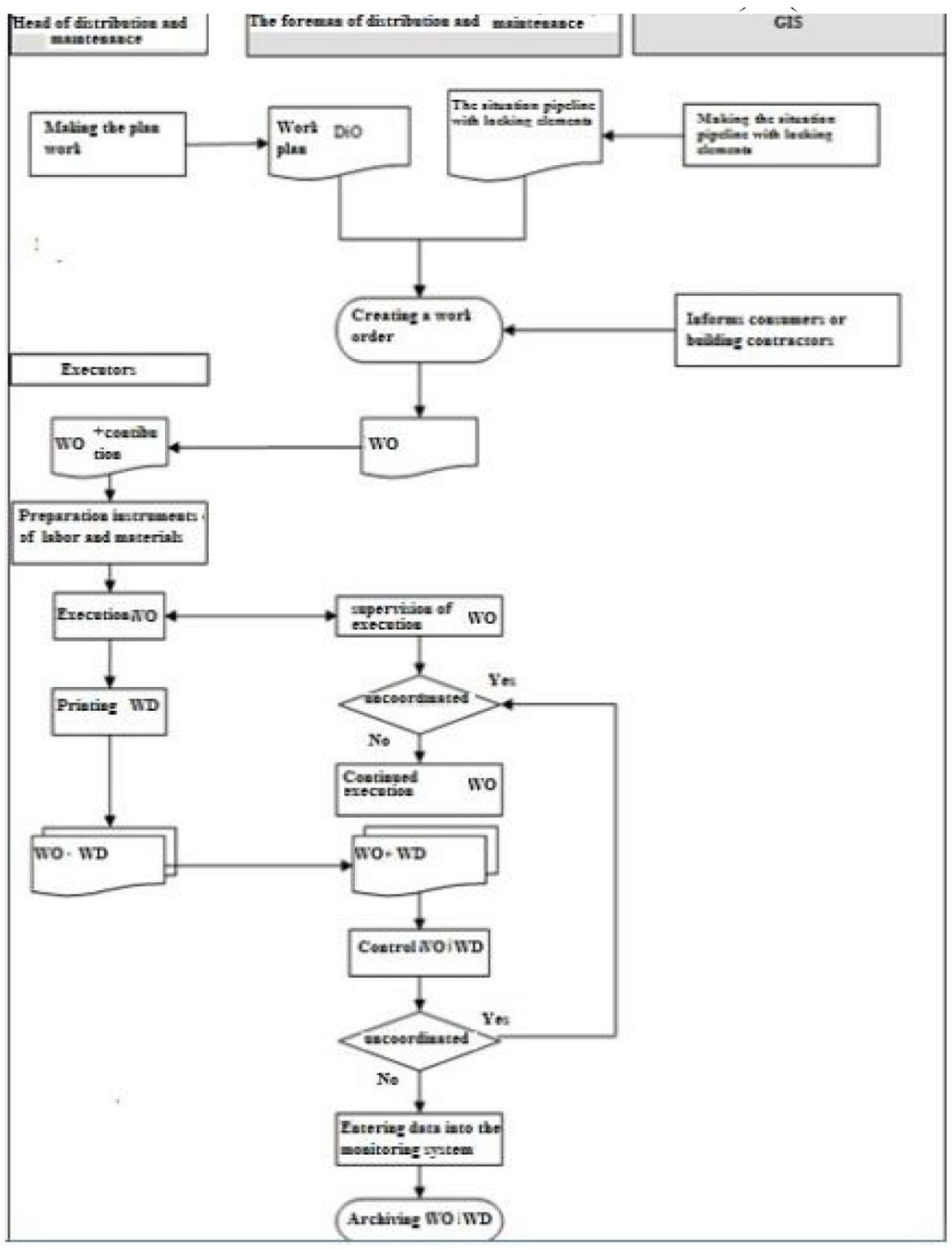

**Fig.8. Diagram shows the flow of activities such as renovation of lines and connectors**

(Source: Development of an author based on interviews of company employees and manual quality (PQ 9001. Prema EN ISO 9001:2000). (12.2.2007))

Fig.8. shows the flow of activities in the repair of lines and connections. Head of distribution and maintenance plan is created works of distribution and maintenance. The manager of distribution and maintenance plan receives distribution and maintenance of works and prepares and publishes work order (with numerous contributions received doers). After bailiffs carry out work order, the manager supervises the works, and makes a decision. Decision determines whether to continue the work developed further defined work order and that it is necessary to make corrections or is work performed satisfactorily. The manager receives work order along with the log of work of bailiffs and exercises control and compares them. If the records of documents compliant, the manager enters the data into the monitoring system and daily work and work order archive.

## CONCLUSION

Herewe develop a researchedmodel for integration of information subsystems and wepresent scientific provenarguments for that integration within the public company through observed matrix business technology (Fig.1 and Fig.4).We show how management information system works with functioning of all other information subsystems:for procurement and financial policy,for information system security, technical information subsystem and legal information subsystem. Based on the ERP subsystem we prove important ability how to link all information subsystems (in departments) within the public company. We describe in detail each information subsystem withits main documents flows, like within technical information subsystem (Fig.7 and Fig.8). The researched model shows the role, application and integration of hardware, software (with database), orgware (organization), computer netware(with networks) and lifeware (with human resources). We show possible threats for information system with its organization and with description of methods of protection of information systems (with the processes and methods of restoration of the database). It can be concluded that the security policy of information subsystems was mainly focused on data and information protection as the results of the execution process. On the basis of the matrix business technology it can be seen how integration of technical information subsystem used class of data belonging to the subsystem procurement and sub-sales(of companies for distribution and supply of gas and other energy). The paper describes the timing of (sub) processes, whereterm (path) of the process should be optimized so as to obtain at the output of processing results is adequate. Optimizing running times (sub) process is required on the achievement of planned results at the exit of the (sub) process and the costs of running the (sub) process to be equal to or less while the quality of the output must be higher than it was before. For information subsystems are given objectives to be accomplished. Shown are flow charts of activities on the basis of which can be seen unfolding sequence of activities during the course of intervention, rehabilitation of power lines and connections. A clear diagram of the flow of activities is a precondition for the construction of an ERP system that would support e-business and information that connects subsystems. ERP systems also have application in the companies which are used to support the business. Related information systems are the model presented in the order of stages in the life cycle of service delivery.

★★★